%% file: ms.tex
\documentclass[
twocolumn,
]{ceurart}

\usepackage{graphicx}
\usepackage{subfiles}

\usepackage{todonotes}
\usepackage{booktabs}
\PassOptionsToPackage{hyphens}{url}\usepackage{hyperref}
\usepackage[hyphens]{xurl}

\usepackage{dsfont}
\usepackage{amsmath}

\usepackage{framed}
\usepackage{caption}
\usepackage{subcaption}

\usepackage{xcolor}

\begin{document}

\copyrightyear{2021}
\copyrightclause{Copyright for this paper by its authors.
  Use permitted under Creative Commons License Attribution 4.0
  International (CC BY 4.0).}

\conference{DESIRES 2021 -- 2nd International Conference on Design of Experimental Search \& Information REtrieval Systems, September 15--18, 2021, Padua, Italy}

\title{TAR on Social Media: A Framework for Online Content Moderation}

\author[1]{Eugene Yang}[
email=eugene@ir.cs.georgetown.edu
]
\address[1]{IR Lab, Georgetown University, Washington, DC, USA}

\author[2]{David D. Lewis}[
email=desires2021paper@davelewis.com
]
\address[2]{Reveal Brainspace, Chicago, IL, USA}

\author[3]{Ophir Frieder}[
email=ophir@ir.cs.georgetown.edu
]
\address[3]{IR Lab, Georgetown University, Washington, DC, USA}

\begin{abstract}
Content moderation (removing or limiting the distribution of posts based on their contents) is one tool social networks use to fight problems such as harassment and disinformation. 
Manually screening all content is usually impractical given the scale of social media data, and the need for nuanced human interpretations makes fully automated approaches infeasible. We consider content moderation from the perspective of technology-assisted review~(TAR): a human-in-the-loop active learning approach developed for high recall retrieval problems in civil litigation and other fields.
We show how TAR workflows, and a TAR cost model, can be adapted to the content moderation problem. We then demonstrate on two publicly available content moderation data sets that a TAR workflow can reduce moderation costs by 20\% to 55\% across a variety of conditions. 
\end{abstract}


\maketitle

\subfile{1-intro}

\subfile{2-background}

\subfile{3-TAR}

\subfile{4-experiment}

\subfile{5-results}

\subfile{6-summary}

\bibliography{sample-base}

\end{document}

%% file: 1-intro.tex
\section{Introduction}

Online social networks are powerful platforms for personal communication, community building, and free expression. Unfortunately, they can also be powerful platforms for harassment, disinformation, and perpetration of criminal and terrorist activities. Organizations hosting social networks, such as Facebook, Twitter, Reddit, and others, have deployed a range of techniques to counteract these threats and maintain a safe and respectful environment for their users.%

One such approach is content moderation: removal (hard moderation) or demoting (soft moderation) of policy-violating posts~\cite{duarte2018mixed, gallo2021congress-report}. 
Despite recent progress in machine learning, online content moderation still heavily relies on human reviews~\cite{ruckenstein2019re}.
Facebook's CEO Mark Zuckerberg stated that language nuances could get lost when relying on automated detection approaches, emphasizing the necessities for human judgments.~\footnote{\url{https://www.businessinsider.com/zuckerberg-nuances-content-moderation-ai-misinformation-hearing-2021-3}} Ongoing changes in what is considered inappropriate content complicates the use of machine learning~\cite{cambridge-report}. Policy experts have argued that complete automation of content moderation is socially undesirable regardless of algorithmic accuracy~\cite{gorwa2020algorithmic}. 

It is thus widely believed that both human moderation and automated classification will be required for online content moderation for the foreseeable future~\cite{duarte2018mixed, gorwa2020algorithmic, gillespie2018custodians}. This has meant not just capital investments in machine learning tools for moderation, but also massive ongoing personnel expenses for teams of human reviewers~\cite{fact.mr2020market}.

Surprisingly, the challenge of reducing costs when both machine learning and manual review are necessary has been an active area of interest for almost two decades, but in a completely different area: civil litigation. Electronic discovery (eDiscovery) projects involve teams of attorneys, sometimes billing the equivalent of hundreds of euros per person-hour, seeking to find documents responsive to a legal matter \cite{surguy2018international}. As the volume of electronically produced documents grew, machine learning began to be integrated in eDiscovery workflows in the early 2000s, a history we review elsewhere~\cite{cost-structure-paper}. 

The result in the legal world has been \textit{technology-assisted review} (TAR): human-in-the-loop active learning workflows that prioritize the most important documents for review~\cite{grossman2017TAR, baron2016perspectives}.
One-phase (continuous model refinement) and two-phase (with separate training and deployment phases) TAR workflows are both in use~\cite{cost-structure-paper, tredennick2015tar}. %

Because of the need to find most or all relevant documents, eDiscovery has been referred to as a \textit{high recall review} (HRR) problem~\cite{oard2010evaluation, totalrecall2015, totalrecall2016}. HRR problems also arise in systematic reviews in medicine, sunshine law requests, and other tasks \cite{baron2017dark,marshall2019toward,wallace2010semi}.  Online content moderation is an HRR problem as well, in that a very high proportion of inappropriate content should be identified and removed.

Our contributions in this paper are two-fold. First, we describe how to adapt TAR and its cost-based evaluation framework to the content moderation problem. Second, we test this approach using two publicly available content moderation datasets. Our experiments show substantial cost reductions using the proposed TAR framework over both manual review of unprioritized documents and training of prioritized models on random samples.

%% file: 2-background.tex
\section{Background}

Content moderation on online platforms is a necessity~\cite{bekkers2013social, veglis2014moderation} and has been argued by some to be the defining feature of an online platform~\cite{gillespie2018custodians}. Despite terms of service and community rules on each platform, users produce inappropriate content, particularly 
when anonymous~\cite{langvardt2017regulating}. Inappropriate content includes toxic content such as hate speech~\cite{macavaney2019hate}, offensive content~\cite{toxccin-paper}, and mis / disinformation~\cite{cambridge-report, toxccin-paper}. It also includes content that is inappropriate for legal or commercial reasons, such as potential copyright violations~\cite{gorwa2020algorithmic, holland2014intermediary}. 

The identification of toxic content can require subtle human insight~\cite{ cambridge-report, macavaney2019hate}, both due to attempts at obfuscation by posters, and because the inappropriateness of the content is often tied to its 
cultural, regional, and temporal context~\cite{duarte2018mixed, ruckenstein2019re}. Mis- and disinformation often consists of subtle mixtures of truthful and misleading content that require human common sense inferences and other background knowledge~\cite{cambridge-report, toxccin-paper}.  

Social media organizations have deployed numerous techniques for implementing community policies, including graph- and time-based analyses of communication patterns, user profile information, and others~\cite{halevy2020preserving}. Our focus here, however, is on methods that use the content of a post. 

Content monitoring falls into three categories: manual moderation, text classification, and human-in-the-loop methods. 
The latter two approaches leverage machine learning models and are sometimes collectively referred to as \textit{algorithmic content moderation} in policy research~\cite{gorwa2020algorithmic}.

Manual moderation is the oldest approach, dating back to email mailing lists. It is, however, extremely expensive at the scale of large social networks and suffers potential human biases. Additionally, mental health concerns are an issue for moderators exposed to large volumes of toxic content~\cite{halevy2020preserving, akhtar2020modeling, sap2019risk}.  

The simplest text classification approaches are keyword filters, but these are susceptible to embarrassing mistakes\footnote{\url{https://www.techdirt.com/articles/20200912/11133045288/paypal-blocks-purchases-tardigrade-merchandise-potentially-violating-us-sanctions-laws.shtml}} and countermeasures by content creators. More effective text classification approaches to content moderation are based on supervised machine learning~\cite{pavlopoulos2017deeper, van2018automatic}. Content types that have been addressed include cyberbullying~\cite{van2018automatic, reynolds2011using, schmidt2017survey, wulczyn2017ex}, hate speech \cite{macavaney2019hate, schmidt2017survey, davidson2017automated, djuric2015hate, fortuna2018survey, nobata2016abusive} or offensive language in general \cite{toxccin-paper, zampieri2019predicting, kumar2018aggression, pitsilis2018detecting, sotudeh2020guir, zampieri2019semeval, zampieri2020semeval}. 

However, some moderation judgments are inevitably too subtle for purely automated methods\footnote{\url{https://venturebeat.com/2020/05/23/ai-proves-its-a-poor-substitute-for-human-content-checkers-during-lockdown/}}, particularly when content is generated with the intent of fooling automated systems \cite{duarte2018mixed, halevy2020preserving, binns2017like}. Content that is  recontextualized from the original problematic context, for example, through reposting, screenshotting, and embedding in new contexts  complicates moderation~\cite{gallo2021congress-report}.  Additionally, bias in automated systems can also arise both by  learning from biased labels and from numerous other choices in data preparation and algorithmic settings~\cite{sap2019risk,dixon2018measuring,mehrabi2019survey}. 
Biased models risk further marginalizing and disproportionately censoring groups that already face discrimination~\cite{duarte2018mixed}.
Differences in cultural and regulatory contexts further complicate the  definition of appropriateness, creating another dimension of complexity when deploying automated content moderation~\cite{cambridge-report}. 

Human-in-the-loop approaches, where AI systems actively manage which materials are brought to the attention of human moderators, attempt to address the weaknesses of both approaches while gathering training data to support supervised learning components \cite{halevy2020preserving, link2016human}. 
Filtering mechanisms that proactively present only approved content (pre-moderation) and/or removal mechanisms that passively take down inappropriate ones are used by platforms depending on the intensity~\cite{cambridge-report}. Reviewing protocols could shift from one to the other based on the frequency of violations or during a specific event, such as elections\footnote{\url{https://www.washingtonpost.com/technology/2020/11/07/facebook-groups-election/}}. Regardless of the workflows, the core and arguably the most critical components is reviews. However, the primary research focus of human-in-the-loop content moderation has been on classification algorithm design and bias mitigation, rarely on the investigation of the overall workflow. 

Like content moderation, eDiscovery is a high recall retrieval task applied to large bodies of primarily textual content (typically enterprise documents, email, and chat)~\cite{baron2016perspectives, tredennick2015tar}.  Both fixed data set and streaming task structures have been explored, though the streaming context tends to bursty (e.g., all data from a single person arriving at once) rather than continuous. Since cost minimization is a primary rationale for TAR~\cite{pace2012where}, research on TAR has focused on training regimens and workflows for minimizing the number, or more generally the cost, of documents reviewed~\cite{cost-structure-paper, tredennick2015tar}.  A new TAR approach is typically evaluated for its ability to meet an effectiveness target while minimizing cost or a cost target while maximizing effectiveness~\cite{wallace2010semi, bagdouri2013towards, cormack2015autonomy}. This makes approaches developed for TAR natural to consider for content moderation.

%% file: 3-TAR.tex
\section{Applying TAR to Content Moderation}
\label{sec:apply-tar}

In most TAR applications, at least a few documents of the (usually rare) category of interest are available at the start of
the workflow.  These are used to initialize an iterative pool-based active learning workflow~\cite{settles2009active}.  Reviewed documents are used to train a predictive model, which in turn is used to select further documents based on predicted relevance~\cite{rocchio1971relevance}, uncertainty~\cite{lewis1994sequential}, or composite factors. Workflows may be batch-oriented (mimicking pre-machine learning manual workflows common in the law) or a stream of documents may be presented through an interactive interface with training done in the background.  These active learning workflows have almost completely displaced training from random examples when supervised learning is used in eDiscovery. 

Two workflow styles can be distinguished~\cite{cost-structure-paper}.  In a \textit{one-phase workflow}, iterative review and training simply continues until a stopping rule is triggered~\cite{cormack2015autonomy, cormack2016engineering, qbcb-paper}. 
Stopping may be conditioned on estimated effectiveness (usually recall), cost limits, and other factors~\cite{cormack2016engineering, quantstop-paper, li2020stop}. 
\textit{Two-phase workflows} stop training before review is finished, and deploy the final trained classifier to rank the remaining documents for review. The reviewed documents are typically drawn from the top of the ranking, with the depth in the ranking chosen so that an estimated effectiveness target is reached~\cite{wallace2010semi, bagdouri2013towards}.  Two-phase workflows are favored when labeling of training data needs to be done by more expensive personnel than are necessary for routine review.

The cost of both one- and two-phase TAR workflows can be captured by in a common cost model~\cite{cost-structure-paper}. The model defines the total cost of a one-phase review terminated at a particular point as the cost incurred in reviewing documents to that point, plus a penalty if the desired effectiveness target (e.g.,  a minimum recall value) has not been met. The penalty is simply the cost of continuing on to an optimal second-phase review from that point, i.e. the minimum number of prioritized documents is reviewed to hit the effectiveness target.  For a two-phase workflow, we similarly define total cost to be the cost of the training phase plus the cost of an optimal second phase using the final trained model.  

These costs in both cases are idealizations in that there may be additional cost (e.g. a labeled random sample) to choose a phase two cutoff~cite{cikmpaper}. However, the model allows a wide range of workflows to be compared on a common basis, as well as allowing differential costs for review of positive vs. negative documents, or phase one vs. phase two documents.

While developed for eDiscovery, the above cost model is also a good fit for content moderation. As discussed in the previous section, the human-in-the-loop moderation approaches used in social media are complex, but in the end reduce to some combination of machine-assisted manual decisions (phase one) and automated decisions based on deploying a trained model (phase two). Operational decisions such as flagging and screening all posts from an account or massive reviewing of posts related to certain events~\cite{cambridge-report, gillespie2018custodians} are all results of applying previously trained models, which is also a form of deployment. Also, broadly applying the model to filter the content vastly reduces moderation burden when similar content is rapidly being published on the platform with the risk of falsely removal~\cite{cambridge-report}. 
We claim no optimal for this specific simplified model in evaluating content moderation, but an initial effort for modeling the human-in-the-loop moderation process. 

When applying the model to content moderation, however, we assume uniform review costs for all documents. This seems the best assumption given the short length of texts reviewed and what is known publicly about the cost structure of moderation~\cite{gillespie2018custodians}.

In the next section, we describe our experimental setting for adapting and evaluating TAR for content moderation.

%% file: 4-experiment.tex
\section{Experiment Design}

Here we review the data sets, evaluation metric, and implementation details for our experiment. 

\subsection{Data Sets}

\begin{figure}[t!]

    \begin{subfigure}{\linewidth}
        \begin{framed}
\begin{flushleft}
shut up mind your own business and go f*** some one else over
\end{flushleft}
	    \end{framed}
	    \vspace{-1mm}
  	    \caption{Wikipedia collection. \\}
    \end{subfigure}
    \hfill \\ \hfill \\
    \begin{subfigure}{\linewidth}
        \begin{framed}
\begin{flushleft}
: being in love with a girl you dont even know yours is sadder\\
: f*** off you f***ing c***!
\end{flushleft}
	    \end{framed}
	    \vspace{-1mm}
  	    \caption{ASKfm collection}
    \end{subfigure}
    \caption{Example content in the collections}
    \label{fig:example_doc}
\end{figure}

We used two fully labeled and publicly available content moderation data sets with a focus on inappropriate user-generated content. The Wikipedia personal attack data set~\cite{wulczyn2017ex}
consists of 115,737 Wikipedia discussion comments with labels obtained via crowdsourcing. An example of the comment is presented in Figure~\ref{fig:example_doc}(a) Eight annotators assigned one of five mutually exclusive labels to each document: Recipient Target, Third Party Target, Quotation Attack, Other Attack, and No Attack (our names). We defined three binary classification tasks corresponding to distinguishing Recipient Target, Third Party Target, or Other Attack from all other classes. (Quotation Attack had too low a prevalence.) A fourth binary classification task distinguished the union of all attacks from No Attack. A document was a positive example if 5 or more annotators put it in the positive class. Proportion of the positive class ranged from 13.44\% to 0.18\%.

The ASKfm cyberbullying dataset~\cite{van2018automatic} 
contains 61,232 English utterance/response pairs, each of which we treated as a single document. An example of the conversation is presented in Figure~\ref{fig:example_doc}(b). Linguists annotated both the poster and responder with zero or one of four mutually exclusive cyberbullying roles, as well as annotating the pair as a whole for any combination of 15 types of textual expressions related to cyberbullying. We treated these annotations as defining 23 binary classifications for a pair, with prevalence of the positive examples ranging from 4.63\% to 0.04\%. 

For both data sets we refer to the binary classification tasks as \textit{topics} and the units being classified as \textit{documents}. 
Documents were tokenized by separating at punctuation and whitespace. Each distinct term became a feature. We used \textit{log tf} weighting as the features for the underlying classification model. The value of a feature was 0 if not present, and else $1+log(tf)$, where $tf$ is the number of occurrences of that term in the document.

\subsection{Algorithms and Workflow}

Our experiments simulated a typical TAR workflow. The first training round is a seed set consisting of one random positive example (simulating manual input) and one random negative example. At the end of each round, a logistic regression model was trained and applied to the unlabeled documents. The training batch for the next round was then selected by one of three methods: a random sampling baseline, uncertainty sampling~\cite{lewis1994sequential}, or relevance feedback (top scoring documents) \cite{rocchio1971relevance}. Variants of the latter two are widely used in eDiscovery \cite{cormack2014evaluation}. Labels for the training batch were looked up, the batch was added to the training set, and a new model trained to repeat the cycle. Batches of size 100 and 200 were used and training continued for 80 and 40 iterations respectively, resulting in 8002 coded training documents at the end.  

We implemented the TAR workflow in  \texttt{libact}\footnote{\url{https://github.com/ntucllab/libact}}~\cite{libact}, an open-source framework for active learning experiments. We fit logistic regression models using Vowpal Wabbit\footnote{\url{https://vowpalwabbit.org/}} with default parameter settings. 
Our experiment framework is available on GitHub\footnote{\url{https://github.com/eugene-yang/TAR-Content-Moderation}}.

\subsection{Evaluation}
Our metric was total cost to reach 80\% recall as described in Section~\ref{sec:apply-tar}. This was computed at the end of each training round as the sum of the number of training documents, plus the ideal second phase review cost as a penalty, which is the number of additional top-ranked documents (if any) needed to bring recall up to 80\%. Ranking was based on sorting the non-training documents by probability of relevance using the most recent trained model. Note that we experimented with 80\% recall as an example. However, the TAR workflow is capable of running with arbitrary recall target, such as 95\% for systematic review~\cite{wallace2010semi, li2020stop}.

In actual TAR workflows, recall would be estimated from a labeled random sample. Since the cost of this sample would be constant across our experimental conditions we used an oracle for recall instead.

%% file: 5-results.tex
\begin{figure*}
  \includegraphics[width=\linewidth]{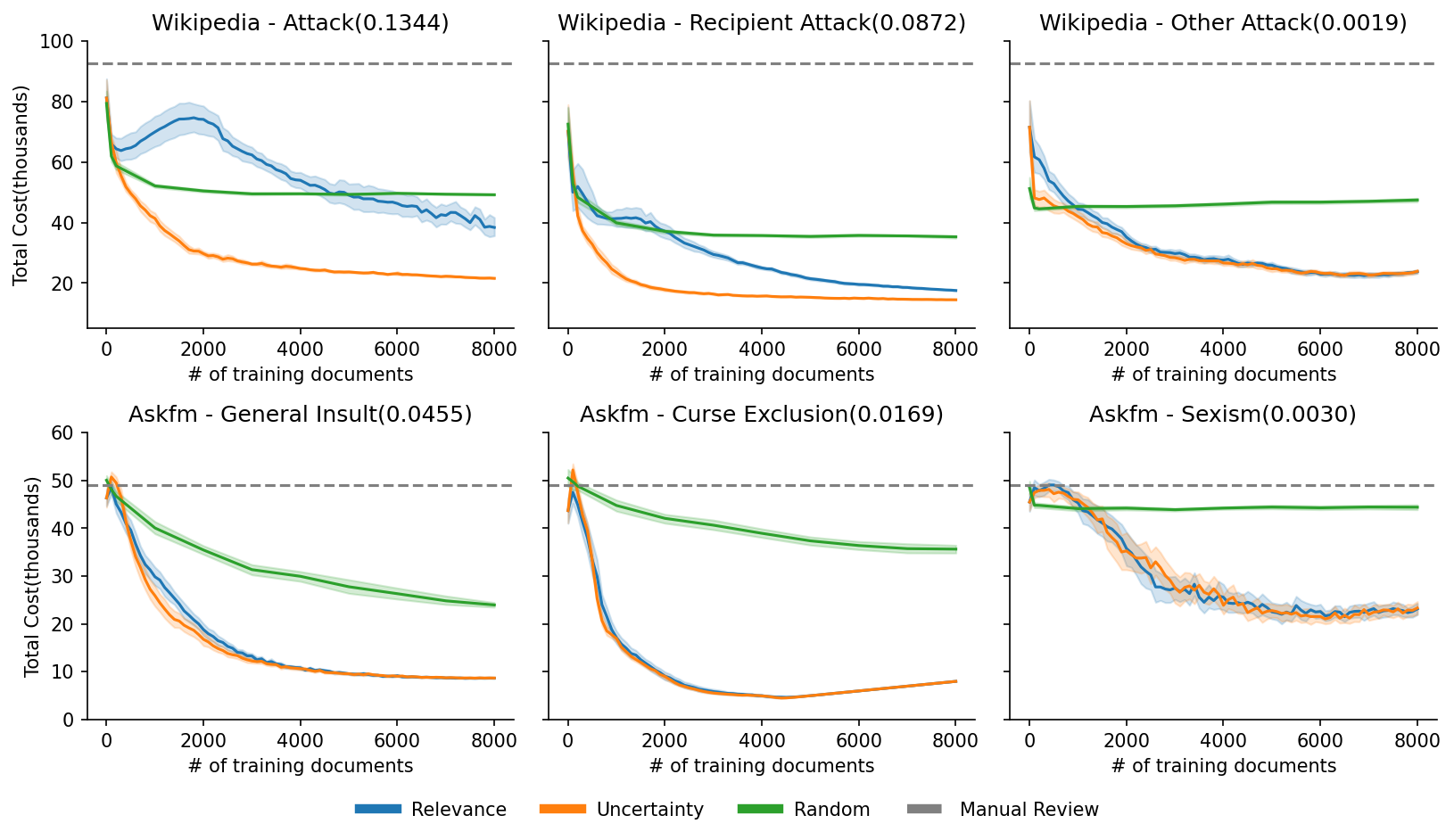}
  \caption{Total cost for TAR alternatives to identify 80\% of positive documents for Wikipedia \textit{Attack}, \textit{Other Attack}, and \textit{Recipient Attack}, and ASKfm \textit{Curse Exclusion}, \textit{General Insult}, and \textit{Sexism} classifications.
  Values are averaged over 20 replicates, and a 99\% confidence interval on costs is shown as shading around each curve. Horizontal line is cost to review a random 80\% of the data set.}
  \label{fig:line_cost}
\end{figure*}

\section{Results and Analysis}

Our core finding was that, as in eDiscovery, active selection of which documents to review reduces costs over random selection. Figure~\ref{fig:line_cost} shows mean cost to reach 80\% recall over 20 replications (different seed sets and random samples) for six representative categories. On all six categories, all TAR workflows within a few iterations beat the baseline of reviewing a random 80\% of the data set (horizontal line labeled Manual Review). 

The Wikipedia \textit{Attack} category is typical of low to moderate prevalence categories ($p=0.1344$). Uncertainty sampling strongly dominates both random sampling (too few positives chosen) and relevance feedback (too many redundant positives chosen for good training). Costs decrease uniformly with additional training. 
We plot 99\% confidence intervals under the assumption that costs are normally distributed across replicates. Costs are not only higher for relevance feedback, but less predictable.    

The ASKfm \textit{Curse Exclusion} ($p=0.0169$) and Wikipedia \textit{Other attack} ($p=0.0019$) category are typical low prevalence categories. Uncertainty sampling and relevance feedback act similarly in such circumstances: even top scoring documents are at best uncertainly positive. Average cost across replicates levels off and starts to increase after 44 iterations for uncertainty sampling and 45 iterations for relevance feedback. This is the point at which additional training no longer pays for itself by improving the ranking of documents. For this category (and typically) this occurs shortly before 80\% recall is reached on the training data alone (iteration 48 for uncertainty sampling and iteration 52 for relevance feedback).     

Task such as the ASKfm \textit{Sexism} category ($p=0.0030$) that deals with nuances in human languages requires more training data to produce a stable classifier. While obtaining training data by random sampling stops reducing the cost after the first iteration, uncertainty sampling and relevance feedback continue to take advantage of additional training data to minimize the cost and become more predictable.  

Note that the general relationship between the prevalence of the task and the cost of reaching a certain recall target using TAR workflows is discussed \citet{cost-structure-paper}.

\begin{table*}
\caption{Total review cost to reach 80\% recall. Values are averaged over all topics for a data set and 20 replicates. Percentages show relative cost reduction over the random sample training baseline. A * indicates that the difference is statistically significant over the random sample training baseline with 99\% confidence by conducting paired t-test with Bonferroni correction. }
\label{tab:total_cost}
\setlength{\tabcolsep}{5pt}
\newcommand{\z}{\phantom{0}}
\renewcommand{\-}{\texttt{-}}
\begin{tabular}{ll|rrr|rrr}
\toprule
      &  {}  & \multicolumn{3}{c|}{ASKfm} &  \multicolumn{3}{c}{Wikipedia}  \\
batch & \# Train & Random  & Relevance & Uncertainty & Random & Relevance & Uncertainty\\
\midrule
100 & 202  &  47685.53 & *49833.73 (\-4.50) & *50273.21 (\-5.43) &  52948.45 &   *60751.69 (\-14.74) &  52210.00 (\z1.39) \\
    & 1002 &  46327.93 & *43329.31 (\z6.47) & *42723.12 (\z7.78) &  49010.71 &   52931.28 (\z\-8.00) &  *39879.78 (18.63) \\
    & 2002 &  45139.15 &  *38179.79 (15.42) &  *37938.19 (15.95) &  47805.25 &   46673.34 (\z\z2.37) &  *29387.06 (38.53) \\
    & 3002 &  44148.28 &  *34909.72 (20.93) &  *34719.50 (21.36) &  47065.66 &   *38964.91 (\z17.21) &  *25676.82 (45.44) \\
    & 4002 &  43731.25 &  *33439.69 (23.53) &  *32795.05 (25.01) &  47234.75 &   *34408.14 (\z27.16) &  *24202.29 (48.76) \\
    & 5002 &  43469.91 &  *32261.33 (25.78) &  *31957.57 (26.48) &  47125.79 &   *31267.88 (\z33.65) &  *22746.94 (51.73) \\
    & 6002 &  42973.85 &  *31767.73 (26.08) &  *31384.51 (26.97) &  47300.02 &   *28945.59 (\z38.80) &  *21922.42 (53.65) \\
    & 7002 &  42563.09 &  *30567.00 (28.18) &  *30502.95 (28.33) &  47086.42 &   *27356.89 (\z41.90) &  *21301.92 (54.76) \\
    & 8002 &  42385.43 &  *30708.85 (27.55) &  *30441.77 (28.18) &  47106.34 &   *25949.51 (\z44.91) &  *21144.28 (55.11) \\
\midrule
200 & 202  &  47685.53 & *49302.36 (\-3.39) & *49339.93 (\-3.47) &  52948.45 &   *58866.41 (\-11.18) &  55747.35 (\-5.29) \\
    & 1002 &  46327.93 &  45014.51 (\z2.84) &  44733.10 (\z3.44) &  49010.71 &   *55302.14 (\-12.84) &  *42896.71 (12.47) \\
    & 2002 &  45139.15 &  *40473.12 (10.34) &  *39894.98 (11.62) &  47805.25 &   49968.88 (\z\-4.53) &  *33981.56 (28.92) \\
    & 3002 &  44148.28 &  *37050.02 (16.08) &  *36902.63 (16.41) &  47065.66 &   42521.55 (\z\z9.65) &  *28332.55 (39.80) \\
    & 4002 &  43731.25 &  *35310.13 (19.26) &  *34888.22 (20.22) &  47234.75 &   *37492.98 (\z20.62) &  *25667.95 (45.66) \\
    & 5002 &  43469.91 &  *33690.33 (22.50) &  *33519.15 (22.89) &  47125.79 &   *34933.90 (\z25.87) &  *24070.44 (48.92) \\
    & 6002 &  42973.85 &  *32425.25 (24.55) &  *32612.13 (24.11) &  47300.02 &   *33004.90 (\z30.22) &  *22839.39 (51.71) \\
    & 7002 &  42563.09 &  *31488.77 (26.02) &  *31813.08 (25.26) &  47086.42 &   *31664.04 (\z32.75) &  *22084.88 (53.10) \\
    & 8002 &  42385.43 &  *31198.75 (26.39) &  *31171.80 (26.46) &  47106.34 &   *29346.76 (\z37.70) &  *21837.84 (53.64) \\
\bottomrule
\end{tabular}

\end{table*}

Table~\ref{tab:total_cost} looks more broadly at the two datasets, averaging costs both over all topics and over 20 replicate runs for each topic for batch sizes of both 100 and 200 . By 20 iterations with batch size of 100 (2002 training documents), TAR workflows with both relevance feedback and uncertainty sampling significantly reduce costs versus TAR with random sampling.  
(Significance is based on paired t-tests assuming non-identical variances and making a Bonferroni correction for 72 tests.) 
All three TAR methods in turn dominate reviewing a random 80\% of the dataset, which costs 92,590 for Wikipedia and 90,958 for ASKfm.

The improvement over cost plateaued after the training sets reached 5000 documents for ASKfm but continue for Wikipedia. Categories in Wikipedia~($p=0.1344$ to $0.0018$) are generally more frequent comparing to ASKfm~($p=0.0463$ to $0.001$), providing more advantage for training to identify more positive documents. Larger batch size slightly reduce the improvement as the underlying classifiers are retrained less frequently. In practice, the sizes are depending on the cost structure of reviewing and specific workflows in each organization. However, as the classifiers are frequently updated with more coded documents, the total cost would be reduced over the iterations.

\begin{figure*}
  \includegraphics[width=\linewidth]{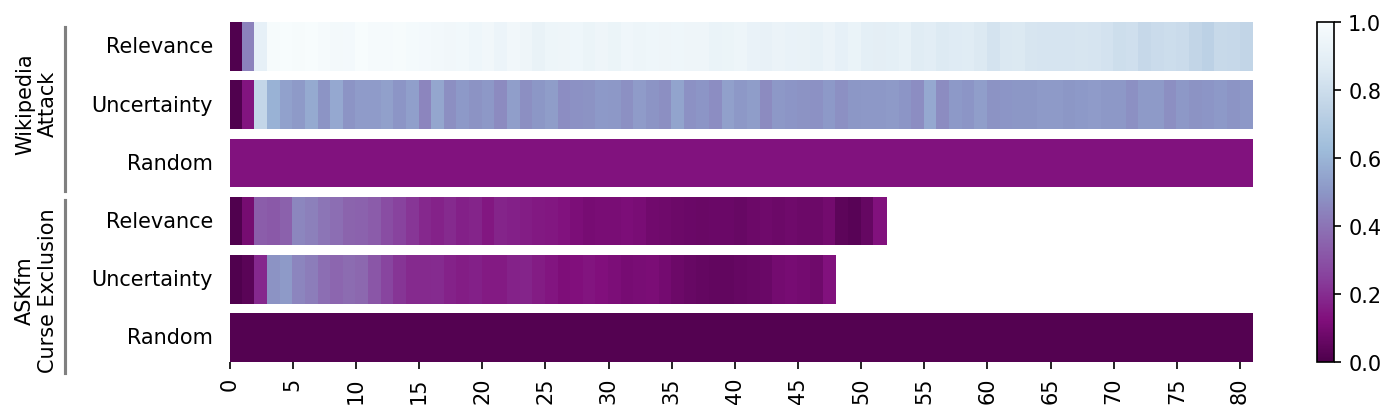}
   \caption{Precision in each batch for TAR workflows on Wikipedia \textit{Attack}($p=0.1344$) and ASKfm \textit{Curse Exclusion}($p=0.0169$) classifications. The x-axis shows the iteration number. A lighter color in an iteration block indicates higher precision.}
  \label{fig:heatmaps}

\end{figure*}

Besides the overall cost reduction, Figure~\ref{fig:heatmaps} shows a heatmap of mean precision across 20 replicates for batches 1 to 81 with batch size of 100, to give insight into the moderator experience of TAR workflows. Precision for relevance feedback starts high and declines very gradually. Uncertainty sampling maintains relatively constant precision. For the very low prevalence category Curse Exclusion we cut off the heatmap at 52 iterations for relevance feedback and 48 iterations for uncertainty sampling since on average 80\% recall is obtained on training data alone by those iterations. For both categories, even applying uncertainty sampling that is intended to improve the quality of the classifier improves the batch precision over the random sampling be a significant amount.

%% file: 6-summary.tex
\section{Summary and Future Work}

Our results suggest that TAR workflows developed for legal review tasks may substantially reduce costs for content moderation tasks. Other legal workflow techniques, such as routing near duplicates and conversational threads in batches to the same reviewer, may be worth testing as well. 

This preliminary experiment omitted complexities that should be explored in more detailed studies. Both content moderation and legal cases involve (at different time scales) streaming collection of data, and concomitant constraints on the time available to make a review decision. Batching and prioritization must reflect these constraints. Moderation in addition must deal with temporal variation in both textual content and the definitions of sensitive content, as well as scaling across many languages and cultures. As litigation and investigations become more international, these challenges may be faced in the law as well, providing opportunity for the legal and moderation fields to learn from each other.